\documentstyle[psfig]{mn}
\input psfig.tex

\def\H0{{\it H}$_0$}
\def\Ms{{\it M}$_\odot$}

\def\q0{{\it q}$_0$}

\def\Ms{{\it M}$_\odot$}

\title[X-ray jets in 3C~9] 
{Chandra reveals a double-sided X-ray jet in the quasar 3C9 at $z=2.012$}  
\author[A.C. Fabian, A. Celotti \& R.M. Johnstone] 
{\parbox[]{6.5in} {A.C. Fabian$^1$, A. Celotti$^2$ and R.M. Johnstone$^1$}\\ \\
$^1$Institute of Astronomy, Madingley Road, Cambridge CB3 0HA\\
$^2$SISSA, via Beirut, 2-4, 34014 Trieste, Italy\\ }

\date{}

\begin{document}

\maketitle

\begin{abstract}
A Chandra observation of the radio-loud quasar 3C9 at redshift
$z=2.012$ has revealed extended X-ray emission coincident with the
radio jet. Of particular interest is the appearance of both jet and
counterjet, which argues against the X-ray emission being highly
beamed. We present the properties of the jets and discuss possible
scenarios for the X-ray emission, contrasting them with the single
sided jet found in PKS\,0637-752, which has a similar X-ray
luminosity. The jet X-ray emission in 3C9 is likely to be due to
either nonthermal emission from a sheath with bulk Lorentz factor less
than 1.5, or thermal emission from shocked cold gas surrounding the
quasar. The thermal possibility implies a high mass for the cold gas
unless it is highly clumped.
\end{abstract}


\section{Introduction}

While the detection of X-ray emission coincident with radio jet
structures is not unprecedented (e.g. Cen A, Schreier et al 1979; M87,
Biretta, Stern \& Harris, 1991; 3C273, Harris \& Stern 1987; R\"oser
et al. 2000), the $Chandra$ X-ray Observatory (CXO) is providing 
an increasing quantity of high spatial and spectral resolution
data on jetted structures associated with Active Galactic Nuclei
(AGN).

The first Chandra observation of a radio--loud quasar, PKS~0637--752,
showed the presence of X-rays extending for $\sim$ 10 arcsec (Chartas
et al. 2000; Schwartz et al. 2000), having close spatial overlap with
the known radio structure.  Since then several jets observed with
Chandra have shown associated X-ray emission (e.g. Pictor A, see
Sambruna et al. 2002; and Worrall, Birkenshaw \& Hardcastle 2001;
Pesce et al 2001; Siemiginowska et al 2002; Hardcastle et al 2002). It
is important to stress that such features have been detected in both
radio--galaxies and blazars (i.e. objects observed at a small angle
with respect to the inner jet direction).

When multifrequency data are available (radio and optical images)
interesting clues on the properties of the emitting plasma are
obtained, such as different trends with distance from the core in
different bands (an intensity decrease is often seen with X-rays
whereas the radio often shows the opposite trend); displacements in
the peak emission and different spatial extensions are sometimes
visible in the images in different bands in the detailed comparison of
individual knot emission.  Broad band information on spatially
resolved structures have also been studied: the local broad band
spectral energy distributions are often not straightforward to
interpret, since in several cases the level of the X-ray emission is
higher than what simple models predict (e.g. Chartas et al. 2000;
Schwartz et al. 2000, R\"oser et al. 2000, Harris \& Krawczynski 2002).

Models involving synchrotron, inverse Compton scattering of the
synchrotron, Cosmic Microwave Background (CMB) and nuclear hidden
blazar radiation to produce X-rays on such scales have been proposed
(e.g. Celotti, Ghisellini \& Chiaberge 2001) as well as models
invoking synchrotron radiation from ultrarelativistic protons
(Aharonian 2000). The most satisfactory model, involving inverse
Compton scattering on the CMB, remarkably requires the presence of
relativistic bulk flows on scale of 100 kpc. Such an emission process
also implies that the X-ray surface brightness of jets is constant
with redshift, as pointed out by Schwartz (2001). Furthermore,
structures in the jet velocity field, typically the presence of a
relativistic core and a slower moving outer `layer' -- as already
suggested on several other grounds -- have been proposed to account for
the varied phenomenology of both blazars and radio--galaxies (e.g.
Celotti et al. 2001).

While much consensus of the above general scenario has accumulated,
X-ray structures with looser morphological similarity to the radio
jets have been found and proposed to be of thermal origin. In
particular recent data from the radio galaxy PKS 1138-262 at $z=2.2$
appear to support the view that a significant fraction of the detected
X-ray flux arises from thermal, shock-heated, gas identified with the
means for confinement for the radio emitting plasma on large scales
(Carilli et al. 2002).

The study of multifrequency emission in jets thus provides important
information on the jet physics and jet environment. Within this
framework, we present here the discovery of extended X-ray emission associated
with the quasar 3C~9, at $z=2.012$. Most notably the X-ray structure
extends on both sides of the core. In this paper we present the data
(Section 2) and discuss the likely origin of the emission (Section 3).

\section{Observations and data analysis}

3C~9 is a powerful quasar at redshift $\sim 2$ with extended radio
emission (e.g. Bridle et al. 1994), classified as FR~II. We observed
it with Chandra for 16,235~s (cleaned of background flares) on 2001
June 10. 3C9 was imaged on the back-illuminated ACIS-B3 chip. The data
clearly show extensions to the NW and SE of the nucleus point source
(Fig.~1). The X-ray structure corresponds well to the radio jets
(Fig.~2). Some possible additional extended X-ray emission is also
seen to the W of the nucleus and in a 'spur' above the SE jet.

\begin{figure}
\centerline{\psfig{figure=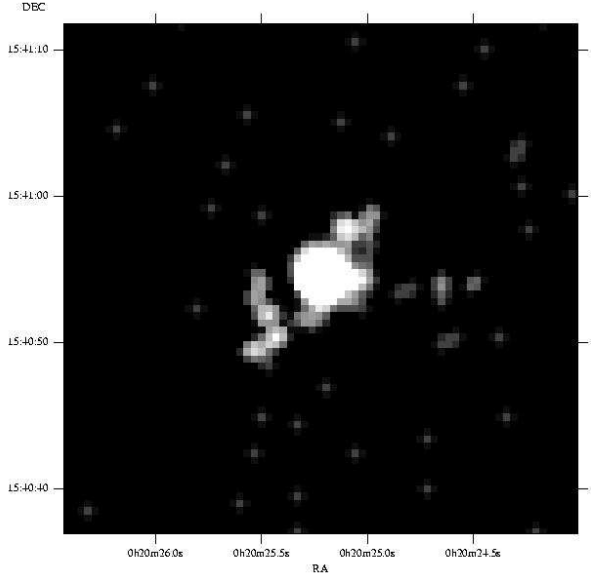,width=0.45\textwidth}} 
\caption{Chandra image of 3C9 in the 0.5--5~keV band. Raw 0.5~arcsec
pixels are used, smoothed by a gaussian of standard deviation one pixel. }
\end{figure}

\begin{figure}
\centerline{\psfig{figure=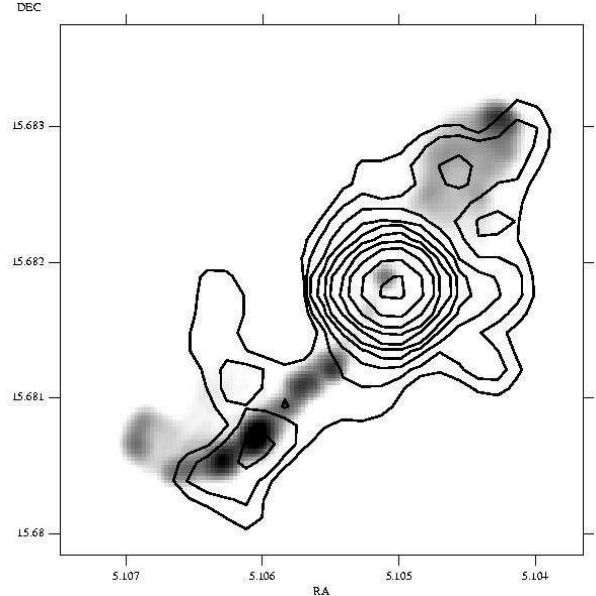,width=0.45\textwidth}} 
\caption{Overlay of 0.5-5 keV X-ray contours (after smoothing the
original 0.5 arcsec binned image with a Gaussian having a standard
deviation of 0.5 arcsec) on the 4.9 GHz VLA radio map (from Bridle et
al 1994).  The contours increase by factors of 1.9 from a level of 0.2
per 0.5 arcsec pixel. A possible $\sim 0.5$ arcsec offset between the
images is within the pointing uncertainty of Chandra, and has not
been corrected for. }
\end{figure}


Excluding the central region the counts in the jet and counterjet
correspond to 18 and 12, respectively. While this of course does not
allow any detailed spectral study, a simple power-law fit to the
spectrum (grouped to a minimum of 10 ct per bin) gives a formal
spectral index $\Gamma_s=1.6 \pm 0.6$ (with a normalization of $K=(2.8
\pm 0.9)\times 10^{-6}$ ph cm$^{-2}$ s$^{-1}$ keV$^{-1}$) assuming the
Galactic value for $N_{H} =4.2 \times 10^{20}$ cm$^{-2}$.
Uncertainities are quoted at the 90 per cent confidence level. The
X-ray flux (0.4--5 keV) amounts to $1.23\times 10^{-14}$ erg cm$^{-2}$
s$^{-1}$ and the corresponding jet $+$ counterjet luminosity is
$\simeq 2.4\times 10^{44}$ erg s$^{-1}$ in the 2--10 keV band (rest
frame) (assuming $H_0= 50$ km s$^{-1}$ Mpc$^{-1}$ and $q_0=0.5$). A
thermal (MEKAL) fit to the spectrum yields a minimum (quasar rest
frame) temperature of 4.1~keV (at the 90 per cent confidence level)
with no constraint on the upper temperature.

The nucleus suffers mild pile-up (at the 4 per cent level). Spectral
fitting of a power-law continuum gives a photon index
$\Gamma_s=1.58\pm0.1$ (similar to that of the extended emission) and a
flux of $1.45\times 10^{-13}$ erg cm$^{-2}$ s$^{-1}$ in the
0.2--3.5~keV band.

\section{Discussion}

High resolution VLA radio imaging of 3C~9 shows two jets (Bridle et al.
1994). As can be seen in Fig.~1 the comparison of the radio and X-ray
images shows that the X-ray extensions roughly coincide with the jet
features, although apparently only in its inner parts before the jet
bends (especially in the S structure).

In order to determine the origin of the large scale X-rays in 3C~9,
assuming it is non--thermal radiation from the radio emitting plasma,
let us first compare its properties with those of PKS~0637-752, at
$z=0.654$.  The total low frequency radio flux (at 408 MHz) of the two
systems are similar, while the extended X-rays are a factor $\sim 10$
lower in 3C~9 (Schwartz et al. 2000). Also the nuclear (blazar-like)
X-ray emission of PKS~0637-752 also exceeds that of 3C~9 by a similar
factor ($\sim 10-50$, Wilkes et al. 1994). 


This implies that both the X-ray extended jet emission and
(especially) the blazar core in 3C~9 are less prominent, with respect
to the presumably isotropic extended radio, than in PKS~0637-752.  The
differing prominence of the blazar component could be simply ascribed
to a difference in the angle between the (core) jet and the line of
sight, as envisaged by unification schemes.

This is indeed supported by the flat radio spectrum of PKS~0637-752
compared to a steep spectrum in 3C~9 ($\alpha_{r}\sim$1).  Furthermore
for PKS~0637-752 superluminal motion indicates bulk Lorentz factors
(relative to the core jet) $\Gamma_b >$ 17 and an angle between the
jet axis and line of sight $\theta <$ 6 deg (Lovell 2000).  On the
other hand for 3C~9 we can consider another crucial point, namely the
jet/counterjet ratios, in the radio and X-ray bands. They are $\sim$
3.3 in the radio (assuming only the straight part) and in the range
0.9-2.6 in X-rays. These are remarkably similar and imply that on the
extended scales $\Gamma_b <$ 1.5 even for an angle of 80 deg (note
that unification models would suggest 45 deg as a limit for a quasar).
We conclude from these elements that 3C~9 is presumably observed at a
larger viewing angle.

What can we then infer on the dominant radiation mechanism for the
extended X-ray emission?  On the one hand the energy density in the
CMB radiation field is about a factor 10 higher in 3C~9. On the other,
however, the nuclear X-ray luminosities (considered indicative of the
blazar emission) of the two objects is of the same order, differing
only by a factor $\sim 2.5$.  While these two facts appear to suggest
that also in 3C~9 the most likely photon field seed of the inverse
Compton emission is the CMB radiation, if one takes into account the
significant difference in viewing angle and thus of beaming, the
blazar in 3C~9 can consistently be an order of magnitude larger (as seen
at small angles).  Together with the limit on $\Gamma_b$ on the extended
scales, this suggests that the energy density associated with the
nuclear emission might dominate over the energy density in the CMB
field and thus in 3C~9 we probably observe emission from a slow
(jet layer?) component (with respect to the case of PKS~0637-752)
scattering the blazar radiation.  Such a possibility is supported by the
morphology of the emission, as in 3C~9 the X-ray extended structure is
(within the limits of resolution) closer to the nucleus up to $\sim$ 100
kpc (for the whole jet), indeed disappearing from detection after the
jet bending, while in PKS~0637-752 it peaks in the large distance knots,
at several hundred kpc. Note that in this situation of low
relativistic speed we do not expect a large effect, e.g. on the
jet/counterjet ratio, due to the anisotropic scattering of the close
and distant jets.

While the evidence discussed so far might be reasonably convincing, an
alternative view could ascribe the majority of the X-ray flux to
thermal emission, as suggested by Carilli et
al. (2002), from gas heated via shocks produced by the jet propagation
and able to confine the jet itself. Indeed the quality of the
morphological similarity between the radio and X-ray spatial
distribution -- even the evidence of a (spatially coincident) knot (in
the S jet) -- cannot distinguish between these possibilities.
Similarly the spectral analysis cannot constrain the nature of the
emission.  Note also that the jet/counterjet
ratio, in principle a powerful diagnostic for discriminating between
the thermal and non-thermal scenarios, is consistent (due to the
paucity of counts) with a similar, isotropic emission from the two jet
sides.

Additional information for X-ray emission models is the extended Lyman
$\alpha$ nebula around 3C9 found by Heckman et al (1991), with a
tentative detection in Lyman $\alpha$ of a structure coincident with
brighter part of the SE jet. This means that there is extensive
ionized gas around the quasar, possibly at high gas pressure, as
inferred by the redshifted H$\alpha$ and [OIII] observations reported
by Wilman, Johnstone \& Crawford (2000). If the extended X-ray
emission is thermal bremsstrahlung then, assuming an emission
temperature of 10~keV for the shocked gas and a volume of $10\times
10\times 50$~kpc$^3$, the electron density is about 0.5~cm$^{-3}$
meaning a pre-shock density of $\sim0.1$~cm$^{-3}$. This is high if
the pre-shock temperature is 1~keV or more, comparable only to that
seen in the centres of strong cooling flows in nearby rich clusters.
The Chandra data are inconsistent with the extensive intracluster
medium of a rich cluster around the quasar. (Ignoring the weak
temperature dependence, the emissivity of the pre-shock gas is $1/16$
of that post-shock, so a volume 16 times that of the jet should have a
similar flux which for a cylinder corresponds to a radius of about 4
arcsec; the data at this radius are however entirely consistent with
the background.) The shock temperature must exceed about
2000~km~s$^{-1}$ to agree with our lower limit on the gas temperature.
The total mass of shocked gas (assuming unit filling factor $f$ of the
volume assumed above) is about $10^{11}$~\Ms.  Presumably there is
much more gas beyond the region now shocked which occupies only a few
per cent of the volume within the radius of the jet, raising the total
gas mass to above $10^{12}$\Ms. The mass varies as $f^{1/2},$
requiring a very low filling factor (say $f\sim 10^{-3}$ if the mass
is to be significantly reduced.

Only if the gas is cooler can high pre-shock densities be present and
consistent with observations. The difficulty is then accounting for
such an extensive and massive atmosphere of relatively dense gas. It
might have to be in the form of many small dense cold clouds in order
to minimise the mass. Perhaps both the X-ray and Lyman $\alpha$
emission are due to shocks and photoionization of a substantial mass
of cold gas clouds which are remnants of galaxy formation and mergers.




Clearly, it is of great relevance to distinguish between these
scenarios with higher quality data, which will allow the nonthermal
jet and thermal shocked gas possibilities to be distinguished, thereby
constraining either the jet velocity field, dissipation and particle
acceleration processes or the interaction of the jet with its
confining mechanism.

\noindent
\section{\bf Acknowledgments}

We thank Alan Bridle for sending us the radio image of 3C9. The Royal
Society (ACF) and the Italian MIUR (AC) are thanked for financial
support.

\end{document}